 \documentclass[apl,reprint,twocolumn]{revtex4-1}
\usepackage{graphicx}
\usepackage[english]{babel}

\usepackage{color}

\usepackage{amsfonts,amssymb,amsmath}

\usepackage[latin1]{inputenc}

\begin{document}

\preprint{APS/123-QED}

\title{THz emission induced by optical beating in nanometer-length 
field-effect transistors}

\date{\today}
\author{P. Nouvel} \author{J. Torres\email{jeremi.torres@ies.univ-montp2.fr }} \author{H. Marinchio} \author{T. Laurent}\author{ C. Palermo}\author{L. Varani}
\affiliation{Institut d'\'Electronique du Sud-CNRS UMR 5214, Montpellier, France}
\author {P. Shiktorov}\author{ E. Starikov}\author{ V. Gruzinskis}
\affiliation{Semiconductor Physics Institute, A. Gostauto 11, Vilnius, Lithuania}
\author{F. Teppe}
\affiliation{Groupe d'Etude des Semiconducteurs-CNRS UMR 5650, Montpellier, France}
\author{Y. Roelens, A. Shchepetov, S. Bollaert} 
\affiliation{Institut d'Electronique de Microelectronique et de Nanotechnologie - UMR 8520, Villeneuve d'Ascq cedex, France }
\begin{abstract}
Experimental results of direct measurement of resonant monochromatic terahertz 
emission optically excited in InGaAs transistor channels are presented. 
The emission is attributed to two-dimensional plasma waves excited 
by photogeneration of electron-hole pairs in the channel at the 
frequency $f_0$ of the beating of two cw-laser sources. 
The presence of resonances for the radiation emission in the range 
of $f_0\pm 10$ GHz (with $f_0$ from 0.3 up to 0.5 THz) detected by a 
Si-bolometer is found. 
Numerical results support that such a high quality of the emission resonances
can be explained by the approach of an instability
in the transistor channel. 

 \end{abstract}
\maketitle

\section{Introduction}
At the present time wide hopes and expectations of development of 
terahertz (THz) range solid-state devices are connected to the use 
of modern field-effect transistors (FETs) and high electron-mobility 
transistors (HEMTs) and their particular features such as high 
mobility of carriers in the channel, strong nonlinearity of current 
flow along the channel governed by a gate, possibility of tuning 
the frequency of two-dimensional (2D) plasma waves excited in the channel 
by varying the gate potential, $\dots$ 
\cite{Dyakonov:1993sj,Dmitriev1996,Dmitriev1997}. 
These peculiarities of carrier transport in the channel allow 
to use such transistors both for detection and generation of THz radiation. 
Recently, the possibility of both resonant and non-resonant detection of 
THz radiation based on these transistors was verified experimentally 
 even at room temperature \cite{knap:2331,Vaksler2006}.

Another possibility widely discussed in the literature 
\cite{lusakowski:064307,dyakonova:141906,meziani:201108,Otsuji2006} 
is to use FETs/HEMTs as a monochromatic 
source of THz radiation where the generation frequency is
controlled by eigen 2D plasma waves excited in the transistor
channel.
Such a situation can be realized when in the transistor conducting channel
a certain plasma instability develops. 
This last may be 
able to resonantly amplify the amplitudes of ac-currents
responsible for the excitation of eigen 2D plasma waves
which, in turn,  play the role of  source of
electromagnetic radiation emitted by the  transistor
into the surrounding space.
The intensity of such an emitted radiation 
will be proportional to the squared amplitude
of the ac-currents amplified by the instability.
The so-called Dyakonov-Shur \cite{Dyakonov:1993sj} instability represents an example of such an amplification effect. 
The possibility to realize other kinds of 
stream-plasma instabilities
\cite{GruzinskisEuro,Mikhailovsky,pierce,Pierce1948},
as well as instabilities related to hot-carrier phenomena
such as Gunn effect\cite{Ridley,Shur1987,StarikovGunn}, 
optical phonon emission assisted transit-time resonance
\cite{starikovJ.Phys2008}, etc. practically are not discussed.
It should be emphasized that, in the case here considered, 
the physical nature of the instability is not a crucial
factor.
The instability merely must amplify the eigen plasma waves
but does not modify significantly their spectrum.
Usually, such a situation takes place in a pre-threshold
state when the system did not begin 
a sharp transformation due to the instability development yet 
but the preparation for such a transformation has already
started.
A typical scenario of such a preparation is associated with the 
growth of the quality of eigen plasma resonances
approaching the threshold.
Here relaxation processes become very long in time
and they are changed by an exponential growth 
of fluctuations above the threshold.
The detailed physical nature of the instability will determine 
in which state (oscillation frequency, linewidth, resonance quality, etc $\dots$) the system will appear finally 
after crossing the threshold.

The important pecularity of the pre-threshold state is that 
the characteristics of the emitted radiation spectrum 
still depend on the physical nature of the source originating 
the initial current  fluctuations in the transistor channel.
Usually, two kinds of such sources are considered:
the spontaneous (internal) source describing the mechanism
of thermal excitation, and the induced (external) one when
the initial fluctuations are excited from outside by some harmonic 
perturbation at frequency $f$.
In the framework of a quasi-classical interpretation of an
initial source as some generalized force, in both cases  
the formal description of the resulting emission spectrum
is the same.
However, the emission spectra and their intensity
will have essential qualitative differences.

In the case of spontaneous excitation it is generally assumed
that the source is given by the internal thermal fluctuations
appearing in the system (the so-called Langevin forces)
whose spectral intensity either is independent of frequency (classical case)
or has a Planck distribution (quantum case) 
\cite{Landau,Lax,ShiktorovPRB1998,ShiktorovRNC2001}.
With such a broadband source, the emission radiation spectrum
mimics the frequency dependence of the emission response
of the system excited by the Langevin force harmonics at frequency $f$.
In the absence of instability, 
a spontaneous excitation initiates merely a 
broad-band emission noise spectrum containing 
more or less pronounced resonances corresponding to the frequencies
of eigen plasma modes of the system \cite{StarikovJStatMech2009}. 
It is evident that, in this case, the realization of a strong 
monochromaticity of the emission requires a rather high level
of resonant amplification by the instability
of the initial thermally-excited plasma oscillations.
In some situations this effect can be realized 
when the instability threshold is overcome
and the system goes into the regime of periodic
self-oscillations with frequencies corresponding
to eigen plasma modes unperturbed by the instability.
One can expect that in the case of a smooth threshold 
the eigen frequencies persist also above the threshold.
Such a transition through the threshold can be realized
usually in submicron semiconductor structures when 
the free-carrier concentration is not sufficient to lead to the formation of  
particular structures like high-field domains, accumulation layers, etc. \cite{starikovJ.Phys2008,ShiktorovICNF2009}

Recent experiments devoted to the investigation of 
the spontaneous emission of THz radiations from AlGaN/GaN based 
HEMT structures \cite{fatimy:024504} demonstrate the presence of 
a threshold for the appearance of a jump-like increase 
of the emission intensity. The authors have related 
this emission to the Dyakonov-Shur 
instability because the emission frequency coincide with the fundamental 
mode of 2D plasma waves. 
Nevertheless, the observed emission spectra are broadband, 
therefore one can conclude that
the instability development process did not reach in this experiment 
an amplification level sufficiently high to overcome 
the relaxation rate. 

The main difference of the induced excitation mechanism 
of initial fluctuations with respect to the spontaneous one
is that the characteristics of the fluctuations source, 
such as its intensity and frequency spectrum,  are now not
controlled  by internal processes.
Rather, they are driven by external processes and hence,
they can be arbitrarily varied within wide limits.
As a result, the radiation emission will take place only at frequencies
where the external action excites ac-currents 
and the spectral width of an emission line
  will be linked to the frequency spreading 
in the source spectrum.
In turn, the emission intensity dependence on
the external source frequency will be determined by internal 
characteristics of the system such as 
 the set of resonant eigenfrequencies,
the relaxation rates, the amplification effects induced by
the development of instabilities.
It is evident that, due to its nature, 
the induced excitation mechanism allows to realize a rather high
level of monochromaticity of the emitted radiation
that will be determined basically by the frequency stability of
the external source of perturbation.
In this case an interesting possibility is to use   
as external source an optical 
photoexcitation creating periodically in time electron-hole 
pairs in a FET/HEMT channel at the beating frequency of two laser beams 
\cite{Otsuji2006}. 
The essence of the effect is that, under photoexcitation
by two laser beams, it is possible to obtain a  periodic in time
generation of extra electron-hole pairs inside the transistor channel.
If the photo-generated holes can quickly leave the channel,
for instance crossing the gate contact, the concentration of 
the remaining photo-electrons in the channel will play the role
of the external harmonic perturbation source
responsible for the induced excitation of 2D-plasma waves.
It is evident that the monochromaticity level of such a 
harmonic excitation will be determined merely by the stability of the 
beating frequency of the two laser beams.
Unfortunately, a direct experimental 
investigation of THz emission under photoexcitation of 2D plasma 
waves is still absent. 
Nevertheless, at the present time there exist indirect experimental 
proofs evidencing the possibility of realization of such a
high-monochromatic emission.
As shown in \cite{Vaksler2006}, due to nonlinearities of the source-to-drain current
in the saturation region of transistors current-voltage characteristics, it is possible to 
achieve the rectification of induced ac-currents. 
Under  current-driven operation mode, this effect manifests itself as
a resonant behavior of changes of the dc-component of 
the source-to-drain voltage drop
when the exciting frequency coincides with an eigenfrequency of the 
2D plasma waves.
Usually, this  effect -- also called self-detection or
 self-control-- is used for the detection of photoexcited 2D plasma 
waves in the FET/HEMT channels \cite{Otsuji:2004ik,torres06}. 
By taking into account that the emission intensity as well as 
the rectification amplitude are proportional to the squared 
amplitude of photo-excited ac-oscillations and, hence,
that their resonant dependences must be similar,
the rectification effect can also be used for the self-control
of the emitted radiation.

In the present communication, the results of direct measurements of 
emission of THz radiations caused by the photoexcitation of 2D plasma 
waves in InGaAs HEMT channels are presented for the first time. 
A comparison with results of the accompanying self-detection of 
the excitation of 2D plasma waves based on rectification effect is 
performed. 
Experimental results are also compared with the numerical results obtained using a hydrodynamic model.

\section{Experimental configuration}
\begin{figure}[htbp]
 \centering
 \includegraphics[width=0.85\columnwidth]{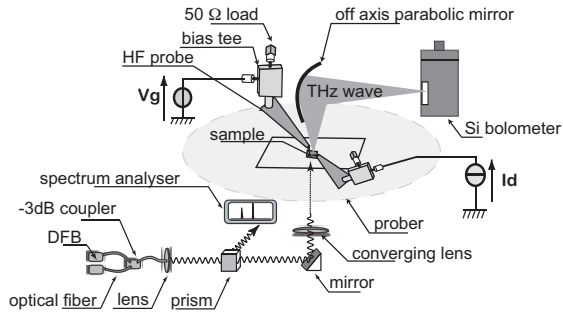}
 \caption{\label{scheme} Experimental configuration scheme. See text for details. }
 \end{figure}

The experimental setup (see Fig. \ref{scheme}) repeats essentially that described in details in ref. \cite{nouvel0x,torres08} and uses two commercially-available DFB continuous-waves-lasers. Their central wavelengths of emission are $\lambda_1$ = 1540.56 nm and $\lambda_2$ = 1542.54 nm and exhibit a short-term spectral width of 2 MHz. Each laser (Power $\approx$ 20 mW) can be tuned over a range of approximately $\pm$ 1.5 nm by changing the temperature (i.e., by varying the current in the Peltier stage). Their mixing, in a 2$\times$2 polarization-maintaining -3 dB coupler, produces a tunable optical beating from in the range 0.3-0.5 THz. The uncertainty in the beating frequency (which is lower than 10 GHz) is estimated by measuring the fluctuations in time of the frequency of each laser with an optical spectrum-analyzer. Additionally, the new setup includes: (i) the system of THz emission detection, and (ii) the sample cooling system. The THz field emitted by the samples is collected by a 2-inches off-axis-parabolic mirror and focused inside a 4K Si-bolometer placed close to the mirror. The temperature is decreased by connecting the sample substrate with thermal transfer ribbons immersed into a nitrogen bath. The temperature of the sample is controlled using a thermocouple. To avoid the formation of ice on the HEMT-top-facet, the experiment is made under an helium flow. Experiments were performed on two HEMTs from InP technology with gate-length values $L_g $ = 50 nm and 400 nm. More detailed description of HEMT layers can be found in ref. \cite{torres06}.
 
\section{Experimental results}
The main experimental results are presented in Figs. \ref{IVcurves} to \ref{photorep} which indicate three specific features of the considered phenomena. 
\subsection{ Static source-to-drain current-voltage characteristics} 
\begin{figure}[htbp]
\centering
\includegraphics[width=0.85\columnwidth]{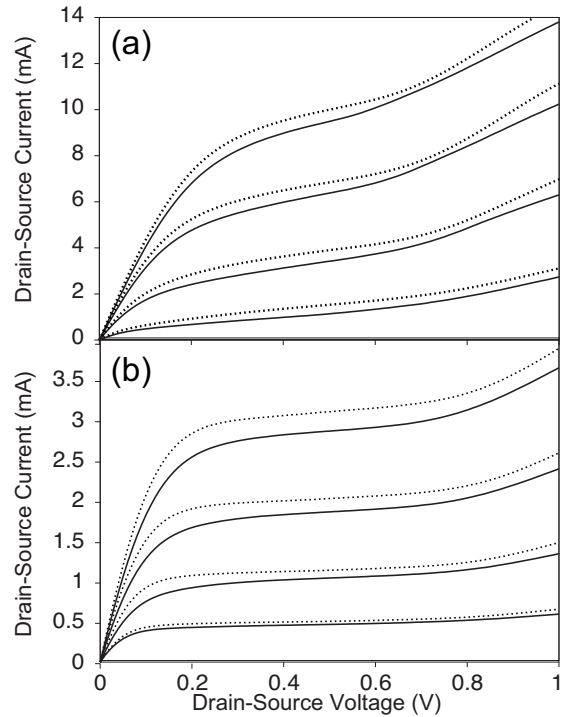}
\caption{\label{IVcurves} Output characteristics at 300K (solid lines) and at 200 K (dotted lines) for a (a)  50 nm and (b) 400 nm gate-length transistor. See text for details.}
\end{figure}

The characteristics of  two HEMT structures, namely:  for (a) a 50 nm gate length HEMT with a threshold voltage $U_{th} \approx -350$\;mV and for (b) a 400 nm gate length HEMT with $U_{th} \approx -400$\;mV are presented in Fig. \ref{IVcurves}, (a) and (b), respectively, for a gate potential $U_g$ varied in the region of about -300 mV to 0 mV with 100 mV step. Solid and dotted lines correspond, respectively, to 300 K and 200 K lattice temperatures. Both HEMTs exhibit the typical current-voltage relation behavior $J_{sd}(U_{sd},U_g)$ of source-to-drain current governed by source-to-drain and gate potentials. With decrease of the lattice temperature from 300 to 200 K, an increase of $J_{sd}$ up to 10-15 $\%$ takes place reaching a maximum value at $U_g\rightarrow 0$~V. Such an increase of  $J_{sd}(U_{sd},U_g)$ agrees well with the change of  free-carrier mobility in the HEMT channel. Estimations of the mobility change based on the IV-relation at low values of $U_{sd}$ and $U_g\approx 0$~V give about a 10-15 $\%$ increase in going from 300 K to 200 K. Note that these IV-relations are independent of the operation regime of the source-to-drain circuit, when: (i) either the voltage-driven operation governed by variations of $U_{sd}$, or (ii) the current-driven operation governed by the current $J_{sd}$ fixed in external circuit are used.

\subsection{Radiation emission from HEMT}
\begin{figure}[htbp]
 \centering
 \includegraphics[width=0.85\columnwidth]{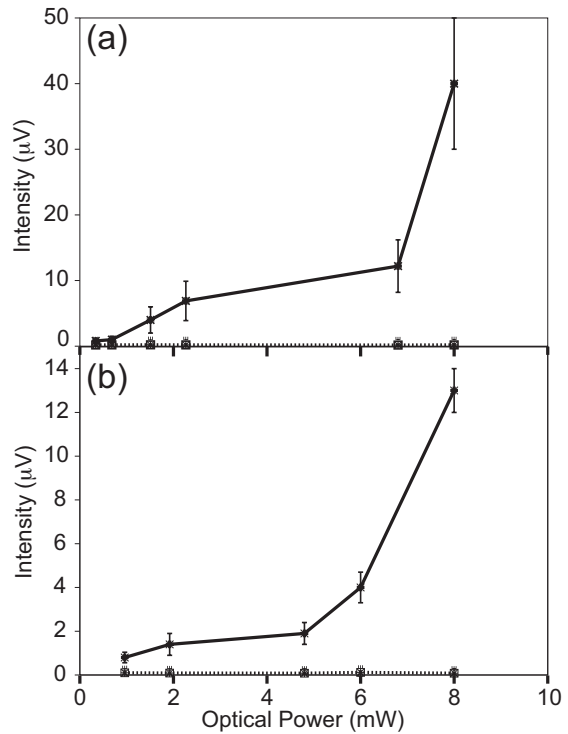}
 \caption{\label{emission-exp-1}  Beam intensity emitted by the HEMTs and measured by a Si-bolometer at 200 K as a function of the power of the optical beating photoexcitation for two transistors with (a) $L_g$ = 50 nm and (b) $L_g$= 400 nm. Solid lines: $f = f_0$ and dotted lines (coinciding practically with the horizontal axis): $f = f_0 \pm$ 10 GHz.}
 \end{figure}

 In the absence of an external excitation the experiment doesn't show 
any manifestation of radiation emission from HEMTs related to the 
self-excitation of plasma waves as predicted in ref. \cite{Dyakonov:1993sj}. 
A sharp resonant growth of the bolometer response was observed only under 
stimulated photoexcitation at $T = 200$ K 
at values of the voltage drop between source-drain contacts 
$U_{sd}\sim 0.4-0.6 \ V$ and at a gate voltage $U_g\ge -0.2\ V$ 
(Fig. \ref{emission-exp-1}). 
In this case, the beating frequency was close (with an accuracy of $\pm$ 10 GHz) to the frequency of the first harmonic of excited plasma waves in the HEMT channel. For the HEMTs with $L_g=$ 50 and 400 nm the resonant frequencies are, respectively, $f_0=0.455$ and $f_0=0.325$ THz (see ref. \cite{nouvel0x} for a more detailed explanation) and they are illustrated by Fig. \ref{emission-exp-1} (a) and (b), respectively. As follows from Fig. \ref{emission-exp-1}, a sharp resonant emission of radiation in the frequency range $f_0\pm 10$ GHz appears. It should be stressed that the bolometer response dependence on the pumping power is non-linear. With the increase of the beating frequency deviation from the resonant frequencies $f_0$ more than 10 GHz, the bolometer response signal went back to the thermal noise level. The minimum frequency step of 10 GHz is dictated by the frequency resolution realized in the beating experiments. The resonant emission takes place only if the transistors are biased under  saturation of the static IV-curves (see Fig. \ref{IVcurves} (a) and (b)), that is at $U_{sd}\ge 0.2$ V. For smaller values of $U_{sd}$, the bolometer response remained within the thermal noise. Let us stress that, under induced excitation, the emission line-width is conditioned by the uncertainty of the excitation source. In these experiments, the uncertainty is not changed when the beating frequency changes and remains at the level of 10 GHz. This allows us to state that the results presented in Fig. \ref{emission-exp-1} evidence that the emission intensity at $T=200$ K has a sharp resonant behavior in the frequency region corresponding to the excitation of the 2D plasma wave fundamental mode. As no emission is measured when the beating frequency varies of more than 10 GHz with respect to the frequency of the fundamental mode, we can state that the resonance Full Width at Half Maximum (FWHM) is not greater than 10 GHz.   
 
 \subsection{Rectification of induced ac currents in HEMT channel} 
 
\begin{figure}[htbp]
\centering
\includegraphics[width=0.85\columnwidth]{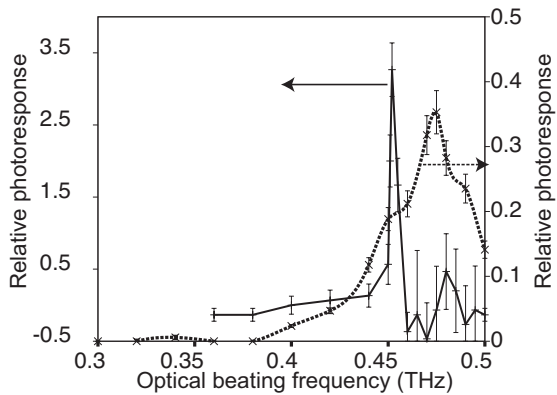}
\caption{\label{photorep} Measured relative photoresponse versus the optical beating frequency for 50 nm gate length HEMT ($U_g = -0.2$ V,  $I_{sd}= 4$ mA at  T= 300 K (dotted line) and at 200 K (solid line). Error-bars are experimental data joined by eye guidelines.}
\end{figure}

Due to the nonlinear character of the current flow in HEMT channels it is possible to realize a self-control of the induced plasma wave intensity by measuring, under fixed drain current regime, the dc component of the source-drain voltage drop $U_{sd}$ as a function of the external signal frequency $f$. As it was shown in \cite{millithaler09,asmontas10}, the resonant dependence on the beating frequency of the amplitudes of both the ac and the dc components of the drain-to-source voltage drop as well as of the ac-current in the channel are identical and are practically characterized by the same FWHM. Therefore, the experimentally obtained resonant behavior of the dc component of the drain-to-source voltage drop (i.e. the photoresponse) will also describe the emission intensity spectrum. Figure \ref{photorep} illustrates the dependence of the relative magnitude of such a photoresponse $R_f$ for the HEMT with $L_g=50$ nm:  
\begin{align}
R_f={U_{sd}(f)  - \overline{ U_{sd}} \over \overline{ U_{sd}}}
\end{align}
where $U_{sd}(f)$ is the source-drain voltage at the beating frequency $f$, while $\overline{ U_{sd}}$ is the source-drain voltage drop at $f=0$. Here dotted and solid lines correspond to measurements performed at 300 and 200 K, respectively and were obtained for the maximum pumping power of 8 mW provided by the laser system. As follows from Fig. \ref{photorep}, the self-detection has a resonant character with a peak at frequencies $f_0=0.475$ and 0.455 THz at 300 and 200 K, respectively. Let us note, that at $T$ = 200 K the resonant frequency of self-detection coincides exactly with the resonant frequency obtained in measurements of the emission intensity (see Fig. \ref{emission-exp-1}). As one can see from Fig. \ref{photorep}, the resonance line-width described by the FWHM sharply decreases from 40 to 5 GHz with temperature decrease from 300 to 200 K. Note, that the estimation of the FWHM at $T$=200 K corresponds to the limiting value determined by the accuracy of the measured beating frequency, therefore the FWHM real value can be even less than the above estimation. An important difference in the results of emission and self-detection is that at 300~K the bolometer does not detect any emission of radiation while the self-detection indicates a resonant photoexcitation of plasma waves at a slightly blue-shifted frequency $f_0=0.475$ THz (with respect to 0.455~THz).  This frequency shift is not considered to be significant due to different experimental conditions when changing the temperature.

\subsection{Discussion}
Summarizing the experimental results, we can state that, in going from 300 
to 200 K: 
(i) the static IV-relations are slightly changed showing 
about 10-20 $\% $ variation of the carrier-mobility in the HEMT channel, 
(ii) both techniques (the former one based on the resonant behavior of
emission and the latter one on the rectification phenomenon) 
allow us to detect photoexcited plasma waves in the HEMT channel 
with eigenfrequencies of 2D plasma waves predicted by the theoretical model 
\cite{marinchioapl09}, 
(iii) detection by using 
the rectification effect shows a photoexcitation of plasma waves at  both 
300 and 200 K with a resonance quality increase of about one order of 
magnitude,
(iv) the direct radiation emission was measured 
only at 200 K while at 300 K it was absent, 
(v) the resonances of emission and rectification responses at 200 K
are characterized by the same value of the FWHM $<$ 10 GHz
which is below the frequency resolution of the setup.

Most of these results 
can be interpreted in the framework of the theory of
quasilinear response to an external perturbation,
which  assumes that 
the photoexcitation regime is linear and, hence, 
the amplitude of the ac-current, $j_d(f)$,
 appearing in the channel at the beating frequency $f$
is proportional to the photoexcitation intensity $P_0$, i.e.
$j_d(f)\sim P_0$.
In the framework of the linear theory,
the experimentally measured signals $W(f)$ 
corresponding to the emission and rectification effects
are related quadratically to the 
beating-induced  ac-current amplitudes as $W(f)\sim |j_d(f)|^2$.
As a consequence:
(i) the emission power must have a quadratic dependence on the
photo-excitation intensity. 
As follows from Fig. 3 the experiment follows approximately this behavior.
(ii) For both effects (emission and rectification)
the dependence of the measured signal amplitudes  
on the beating frequency must be similar. 
The experimental results (see Figs. 3 and 4 
as well as their comments in the text
at 200 K) satisfy this condition.
In both cases the FWHM of the observed 
spectrum does not exceed 10 GHz.
Probably, this value corresponds to the maximum frequency resolution 
of the experimental setup.

As it was noticed above, at 300 K the emission response was
not observed. 
This also can be explained in the framework of the linear response.
Following our estimations such a behavior is connected firstly to a not 
sufficient sensitivity of the setup
used in the experiment for the detection of the emitted  radiation.
Indeed, since 
the frequency dependence of the emission and rectification responses
must have similar behaviors,
from Fig. 4 we can conclude that the emission power 
must decrease of about one order of magnitude in going from
200 to 300 K.
For instance, by taking into account the sensitivity of the detection setup,
at a level of $\sim 15\%$ with respect to the signal
magnitude at 200 K, as it takes place in the experiment (see, e.g.,
vertical bars in Fig. 3), 
the emission response at 300 K cannot be extracted
from the noise level.

At the same time,
the origin of the sharp narrowing of the emission resonance 
with decrease of the lattice temperature cannot  be explained
in the framework of the linear response theory.
Therefore the following question appears :
why a rather low decrease (about 10-20$\%$)
of the velocity relaxation rate (i.e. of the mobility) 
in the HEMT channel   in going from 300 to 200 K
leads to a considerable  enhancement of the resonance quality
(about one order of magnitude) in measurements
of the emitted and rectificated signals?
In the next section,
 we will show that such a behavior
can be explained by the presence of a plasma instability
in the stable situation when the amplification induced by the
instability is still not sufficient to compensate 
the damping caused by the velocity relaxation rate.

\section{Theoretical consideration}
In this section we will analyze in detail 
the experimentally observed sharp narrowing
of the plasma resonance lines in the emission and rectification response
spectra when the FET/HEMT temperature decreases.  
Let us restrict ourselves to the simplest analytical model
which allows, from one hand, 
to describe qualitatively and quantitatively the experimentally observed resonances of 2D plasma
waves excited in FET/HEMT channels \cite{Vaksler2006},
 and from the other hand,
to consider the development of Dyakonov-Shur instability of such waves
which can appear due to the presence of carrier stream along 
the channel \cite{Dmitriev1997,Vaksler2006}.
\subsection{Theoretical model}
We assume that 
carrier transport in the transistor 
can take place only along the conductive channel from
source to drain contacts, i.e. carrier flow in
the transverse direction with respect to the channel
is absent.
This allows us to describe carrier transport 
as a one-dimensional (1D)
process by using simple hydrodynamic (HD) equations [A2]
$$
{\partial n \over \partial t}+
{\partial nv \over \partial x}=G(t)
\eqno(2)
$$
$$
{\partial v \over \partial t}+
{\partial \over \partial x } [{v^2\over 2}
+{e\over m^*} \varphi ]
+ e\nu D{\partial n \over \partial x}+v\nu =0
\eqno(3)
$$
where $n$ and $v$ are the concentration and velocity of electrons
in the channel, respectively,
$\nu$ is the velocity relaxation rate,
$m^*$ the electron mass, 
$D$ the longitudinal diffusion coefficient,
and $G(t)$ is the electron photo-excitation generation term.

The self-consistent potential $\varphi(x)$ inside the channel is
described by a 1D approximation of the 2D Poisson equation [A2]:
$$
\varepsilon_c
{\partial^2 \over \partial x^2}\varphi
+\varepsilon_s{{U_g-\varphi}\over d(x)\delta}=
{e\over \varepsilon_0}[n(x)-N_D(x)]
\eqno(4)
$$
where $\delta$ is the channel width,
$U_g$ the gate potential,
$N_D$ the effective donor concentration in the channel,
$d(x)$ the effective gate-to-channel distance.
A dependence of $d(x)$ on the coordinate in the channel
allows us to describe in the framework of Eq. (4) 
both gated regions where $d(x)$ has certain finite value
and ungated regions where $d(x)\rightarrow \infty$ 
is supposed to tend to infinity.

It is worthwhile to stress limitations of the model based on Eqs. (2) to
(4): 

(i) Only one type of photo-generated carriers, i.e. electrons,
are taken into account. The photo-generated holes contribution to
the space charge and current along the channel is omitted
by supposing that holes, just after photo-excitation, are practically
immediately removed from the channel through the gate.
The time dependence of the optical generation rate
is taken in the form:
$$
G(t)=P_0[1+cos(2\pi f t)]
\eqno(5)$$
with $f$ being the beating frequency and $P_0$
the   photo-generation intensity.

(ii) Hot-carrier effects are not taken into account since
the electron velocity relaxation rate $\nu$ is supposed 
to be constant along the channel and to be not  
dependent of the electrons local energy.
This assumption allows us to consider the development
in  FET/HEMT channels of mono-stream plasma instabilities only.

Under voltage driven operation
the system of Eqs. (2) to (4) is closed,
thus allowing us to calculate various steady-state and 
transient characteristics. 
In this case as boundary conditions for Eq. (4) 
we have used potentials $\varphi(0)$ (usually equal to zero)
and $\varphi(L)$ given in points $x=0$ and $x=L$, respectively,
where $L$ is the channel total length. 

Under current-driven operation
one fixes the total current at the drain contact 
$j_{tot}(L)=const$.
Since the total current consists of the sum of conduction
and displacement currents, the requirement to keep
constant the total current can be reformulated
in terms of an additional equation for the time-evolution 
of the electric field at the drain contact as:
$$
{\partial \over \partial t}E(L,t)=
{1\over \varepsilon_c \varepsilon_0}
\left[
j_{tot}-en(L,t)v(L,t) 
\right]
\eqno(6)
$$
In this case Eq. (6) is solved in parallel with Eqs. (2) and (3).
The boundary condition for the Poisson equation at the source
contact remains the same ($\varphi(0)$) while the drain boundary
condition is rewritten in terms of the electric field $E(L,t)$. 

In our calculations we assumed
that  the lattice temperature is of about T=200 K 
and that the corresponding value of the velocity relaxation rate is of about
$\nu=2 \ (ps)^{-1}$.
Such estimations of the relaxation rate of the electron velocity
in the channel near thermal equilibrium are 
based on Hall-measurements of the  investigated here HEMTs
and on MC simulations of bulk $In_{0.53}Ga_{0.47}As$ 
at different temperatures.
The diffusion coefficient is calculated from the 
low-field mobility $\mu$ by using Einstein relation
$D=kT\mu/e$.

\subsection{Numerical results}
\begin{figure}[htbp]
\centering
\includegraphics[width=0.9\columnwidth]{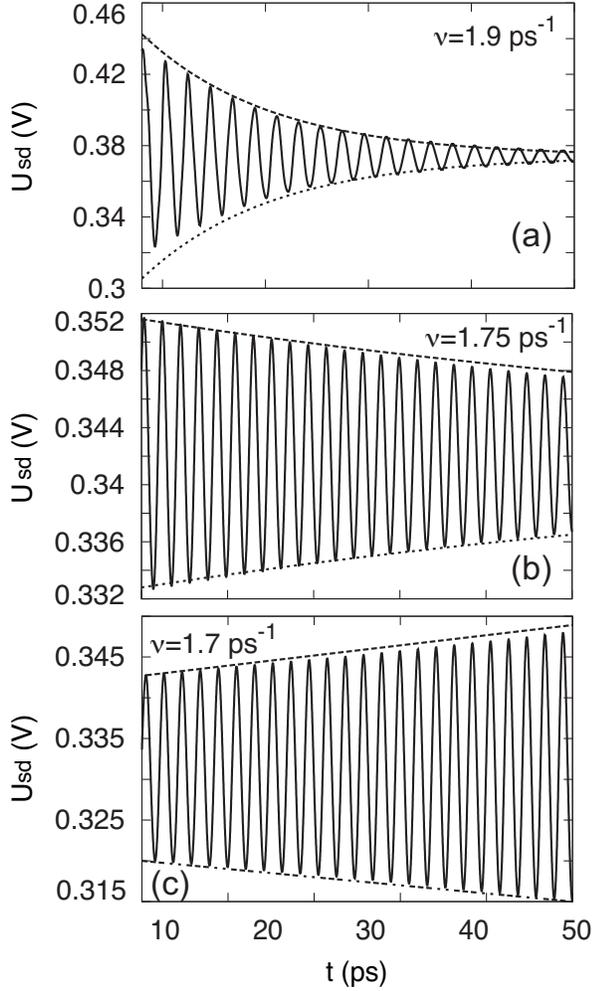}
\caption{\label{oscill} Calculated source-drain voltages as functions of time for the reported electron velocity relaxation rates $\nu$= 1.9 (a), 1.75 (b) and 1.7 (c) ps$^{-1}$. Dotted lines represent the envelopes of the time dependent voltages.}
\end{figure}

As it was already mentioned,
the above model can reproduce the classical Dyakonov-Shur
instability of 2D plasma waves caused by mono-stream of carriers
under the gate. 
Figures \ref{oscill} show the time dependences of the source-to-drain voltages corresponding 
to conditions of progressive development 
of the Dyakonov-Shur instability associated with a 
 a small decrease of the electron velocity
relaxation rates: 1.9, 1.75 and 1.7 ps$^{-1}$ respectively. 
The simulated structure and the bias conditions correspond to 
the experimental ones. In these figures, the relaxation process 
to the stationary
state have an oscillating character with a frequency
corresponding to the main harmonics of 2D plasma waves
excited in the gated regions of the channel.
The main difference is related to which stationary state 
the system relaxes.
In the case of $\nu \ge 1.75$~ps$^{-1}$ (Fig. \ref{oscill} (a)),
the relaxation process leads the system to a time-independent stationary state 
through damped oscillations (the usual dissipative process of an oscillator).
The peculiarity of the relaxation process presented in Fig. \ref{oscill} (b) 
is that, in contrast to the previous case, 
a rather small
decrease of the relaxation rate (about 10 $\%$) results in
a considerable decrease of the system relaxation (approximately one order in magnitude).

A further decrease of the velocity relaxation rate,
$\nu < 1.75$~ps$^{-1}$ (Fig. \ref{oscill} (c)),
changes qualitatively the system state to which
the relaxation process tends.
This state is oscillating in time 
with a constant increase of the oscillations amplitude
(self-oscillation regime).
Let us stress that the relaxation processes presented in Figs. \ref{oscill} 
are obtained in the absence of any optical beating and the change of the state to which
the system relaxes is determined only by processes occurring inside the system.

\begin{figure}[htbp]
\centering
\includegraphics[width=0.9\columnwidth]{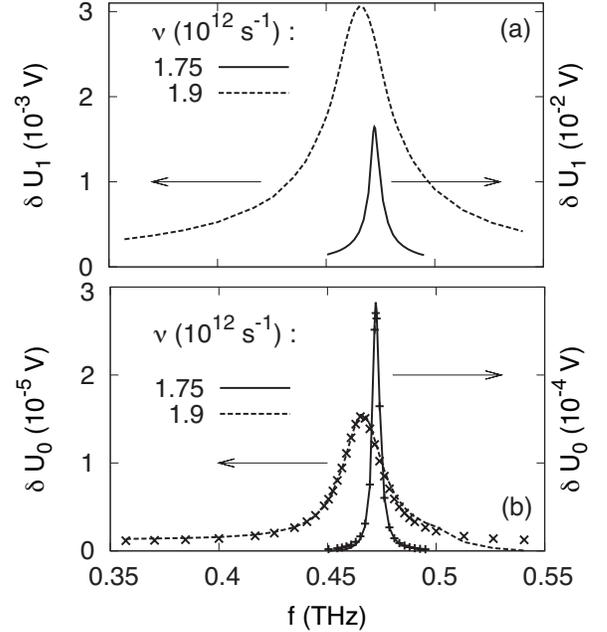}
\caption{\label{simul} Calculated amplitudes of the (a) harmonic  and 
(b) average photoresponses of voltage as functions of the beating 
frequency for the reported values of the velocity relaxation rates.
}
\end{figure}

Figures \ref{simul} (a) and (b) illustrate a typical pre-threshold behavior
of the resonances of ac and dc responses, respectively,
induced by photoexcitation at the beating frequency 
under constant-current operation in the source-drain circuit.
Here, the solid line corresponds to the velocity relaxation rate
$\nu=1.75$~(ps)$^{-1}$ (just before the threshold for the Dyakonov-Shur
self-oscillations to appear in this situation),
 while the dashed line calculated at $\nu=1.9$~(ps)$^{-1}$
illustrates a situation which is still far from self-oscillations.
In this example, the transition into the self-oscillation regime
takes place for  $\nu \lesssim 1.75$~(ps)$^{-1}$
where a clear-cut self-oscillations appear in the simulations (Fig. \ref{oscill} (c)). Symbols in Fig.  \ref{simul} (b) correspond to the square of the ac response
amplitude presented in Fig.  \ref{simul} (a) after normalization to
the dc resonance maximum.
Figure  \ref{simul} (b) fully confirms that
 the rectification response spectrum  $\delta U_0(f)$ 
is originated by a resonant behavior of the ac components 
of the response    
and is related to them by a quadratic dependence.
As follows from Fig.  \ref{simul}, the calculations show 
that, in the pre-threshold regime of the Dyakonov-Shur instability, 
a small variation (about 10$\%$) of the relaxation
rate $\nu$ is able to cause a significant variation 
(about one order of magnitude) 
of the resonance quality of both ac and dc responses 
under photoexcitation. 
These theoretical results are in good agreement with 
the experimentally observed behavior of the 
rectification and emission resonances.
\subsection{Discussion}
The above presented simple HD model demonstrated
that a process like the development of the Dyakonov-Shur instability
can explain and describe 
the experimentally observed 
behavior of the emission and rectification resonances
caused by the photo-excitation of 2D plasma waves in the FET/HEMT
channels.
It is worthwhile to emphasize that the sharp narrowing of
the plasma resonances taking place under stationary conditions just before the transition into the
self-oscillation regime
(obtained here for the Dyakonov-Shur instability)
has a general character and it describes the typical scenario
of a plasma-instability development independently of 
its physical nature.
It should be stressed that
this theoretical model 
can describe only the  stream-plasma instabilities: 
the Dyakonov-Shur instability which was obtained in the modeling 
just belongs to this class. 
Other instabilities related to hot-carrier effects are outside of the present model.

Here it should be stressed that the effect was observed at values of 
the source-drain voltage drop $U_{sd}\sim 0.4-0.6\ V$ 
and at a gate voltage $U_g\ge -0.2\ V$, 
when, in principle, one cannot neglect hot-carrier effects 
such as electron-transfer into upper valleys and  impact ionization.  
In particular, the latter effect explains the increase of the  current at $U_{sd}\ge 0.6 V$ in Fig. 2
Therefore, the presented experimental results on the resonant radiation
emission and rectification effect on the beating frequency
under photo-excitation surely evidence the presence of
an instability  of 2D plasma waves
responsible for the observed effect.
However, the performed experiments cannot elucidate at present the exact microscopic nature of this instability.

\section{Conclusion}

 We have demonstrated both experimentally and theoretically the possibility to obtain emission of THz radiations by exciting a HEMT channel by an optical beating. This emission  appears to be due to 2D plasma oscillations excited by the photogeneration of electron-hole pairs in the channel and is consequently obtained when the beating frequency is tuned to the first eigenfrequency  of 2D plasma waves. In our experiments, such an emission has been observed only at 200 K, when the dynamic behavior of the 2D plasma becomes sufficiently resonant to provide a measurable emission. The spectrum of the emitted radiations is directly related to the spectrum of the excitation, therefore the studied device may be used as a source of monochromatic THz radiation.

\section*{Acknowledgments}Authors wish to thank T. Gonzalez, J. Mateos 
(University of Salamanca, Spain) for useful discussions. This work was partially supported by 
GIS ''TeraLab-Montpellier'' and by grant 
No MIP-87/2010 of Lithuanian Science Council.

\end{document}